\documentclass[aip,
 sd,%
 amsmath,amssymb,
  preprint,%
]{revtex4-1}

\usepackage{graphicx}
\usepackage{dcolumn}
\usepackage{bm}

\usepackage{natbib}
\usepackage{units}
\usepackage{graphicx}
\usepackage{epstopdf}
\usepackage{hyperref}
\usepackage{color}
\usepackage[utf8x]{inputenc}
\usepackage{empheq}
\usepackage{cases}

\DeclareMathOperator{\sech}{sech}

\begin{document}

\preprint{AIP/123-QED}

\title[]{On the electron dynamics during island coalescence in asymmetric magnetic reconnection}

\author{E.~Cazzola}
\email{emanuele.cazzola@wis.leuven.be}
\affiliation{Center for mathematical Plasma Astrophysics, Department of Mathematics, K.U. Leuven (University of Leuven), Celestijnenlaan 200B, B-3001 Leuven, Belgium}

\author{M.~E.~Innocenti}
\email{mariaelena.innocenti@wis.leuven.be}
\affiliation{Center for mathematical Plasma Astrophysics, Department of Mathematics, K.U. Leuven (University of Leuven), Celestijnenlaan 200B, B-3001 Leuven, Belgium}

\author{S.~Markidis}
\email{markidis@pdc.kth.se}
\affiliation{PDC Center for high Performance Computing, KTH Royal Institute of Technology, Teknikringen 14, 10044 Stockholm, Sweden}

\author{M.~Goldman}
\email{martin.goldman@Colorado.edu}
\affiliation{Center for Integrated Plasma Studies, University of Colorado Boulder, Gamow Tower, Boulder, 80309-0390 Colorado, USA}

\author{D.~Newman}
\email{david.newman@colorado.edu}
\affiliation{Center for Integrated Plasma Studies, University of Colorado Boulder, Gamow Tower, Boulder, 80309-0390 Colorado, USA}

\author{G.~Lapenta}
\email{giovanni.lapenta@wis.kuleuven.be}
\affiliation{Center for mathematical Plasma Astrophysics, Department of Mathematics, K.U. Leuven (University of Leuven), Celestijnenlaan 200B, B-3001 Leuven, Belgium}

\date{\today}
             
\begin{abstract}

We present an analysis of the electron dynamics during rapid island merging in asymmetric magnetic
reconnection. We consider a doubly periodic system with two asymmetric transitions. The upper layer is an asymmetric Harris sheet of finite width perturbed initially to promote a single reconnection site. 
The lower layer is  a tangential discontinuity that promotes the formation of many X-points, separated by rapidly merging islands. Across both layers the magnetic field and the density have a  strong jump, but the pressure is held constant.

Our analysis focuses on the consequences of electron energization during island coalescence. We focus first on  the parallel
and perpendicular components of the electron temperature to establish the presence of possible anisotropies and non-gyrotropies. Thanks to the
direct comparison between the two different layers simulated, we can distinguish three main types of behavior characteristic of three different regions of interest. The first type represents the regions 
where traditional asymmetric reconnections take place without involving island merging.
The second type of regions instead show reconnection events between two merging
islands. 
Finally,
the third regions identifies the regions between two diverging island and where typical signature of
reconnection is not observed. Electrons in these latter regions additionally show a flat-top distribution resulting from the saturation of a two-stream instability 
generated by the two interacting electron beams from the two nearest reconnection points.
Finally, the analysis of agyrotropy shows the presence of a
distinct double structure laying all over the lower side facing the higher magnetic field region. This structure becomes quadrupolar in
the proximity of the regions of the third type. 

The distinguishing features found for the three types of regions investigated provide clear indicators to 
the recently launched Magnetospheric Multiscale NASA mission (MMS) for investigating magnetopause reconnection involving multiple islands.

\end{abstract}

\pacs{Valid PACS appear here}
\keywords{Asymmetric magnetic reconnection, asymmetric island merging and coalescence, electrons energetics}

\maketitle

\section{\label{sec:introduction} Introduction}

Magnetic reconnection is a multi-scale non-linear process occurring in plasmas. It is believed to be responsible for explosive events occurring in space which ultimately result in large
release of energy \cite{hoyle1949,priest2007}. Examples are solar flares and 
geomagnetic substorms, as well as processes in laboratory plasma devices, such as, sawtooth disruption occurring in confined fusion plasma.

Kinetic studies of reconnection uncovered important aspects missed by previous reduced models \cite{birn2007}. 
However, the majority of the literature deals with symmetric conditions, relevant for example to the Earth magnetotail \cite{birn2001,shay1998,hesse1999}.

Other interesting reconnecting sites present instead strong
asymmetries in density, temperature and magnetic field. This is the case at the dayside magnetopause, where the shocked solar wind in the
magnetosheath comes in contact with the magnetospheric plasma. In this
region, density drops a factor of ten to thirty between the two sides, while the magnetic field increases around three
times \cite{phan1996,mozer2002,mozer2008themis}.

Simulations under asymmetric conditions yield peculiar results. It
has been already established that plasmoids growing during reconnection tend to
mainly swell toward the weakest magnetic field side due to a lower magnetic resistance from the local field \cite{cassak2007,pritchett2008}.
Further interesting properties are the X-line drift, and
the displacement between the X-line and stagnation point, which leads to an
unusual net plasma flow through the null point. These latter processes are mainly governed by energy and mass fluxes. Due to a magnetic flux
imbalance between the two sides, corresponding to a magnetic pressure $ \propto vB^2$ imbalance, the X-line is always displaced toward the weakest field region. Conversely,
the stagnation point is shifted toward the side with the smallest mass
flux, which is $\propto \rho B$.
Moreover, when a guide field is present, the X-line experiences a lateral
drift caused by the electron diamagnetic velocity
\cite{ugai2000,swisdak2003}. When the relative ion-electron drift
approaches the Alfvén speed, the magnetic reconnection rate is reported to be
strongly reduced or even suppressed \cite{swisdak2003}. 

These properties have been studied either using an MHD approach 
\cite{ugai2000,cassak2007,borovsky2007,cassak2008} or kinetic codes 
\cite{swisdak2003,pritchett2008,pritchett2009,malakit2010}, as well as through
observations \cite{mozer2002,mozer2008themis,mozer2008}. Important scaling analysis between macroscopic
quantities have been derived theoretically by \textcite{cassak2007,cassak2008}, and confirmed with MHD simulations \cite{borovsky2007,cassak2007,cassak2008}, as
well as with kinetic simulations \cite{malakit2010}. Although originally derived from Sweet-Parker model for collisional plasmas, this conclusion still holds for the collisionless plasmas in asymmetric 
configuration, as no
assumptions on the dissipation mechanism were effectively required.

MHD approaches have been a very successful method
in assessing the plasma behavior for large macroscopic scales. A kinetic
approach (e.g. a Particle-in-Cell (PIC)) \cite{hockney1988,birdsall1991,verboncoeur2005}, is however essential to fully understand the physics responsible for the dissipative effects involved in reconnection. 
On the other hand, high resolution kinetic simulations of small scales are very computational demanding.
The kinetic effects on reconnection are well-established to occur in two nested regions, namely in the outer ion diffusion region (as thick as the ion skin depth $\sim d_i$) and the inner 
reconnection site (also called the electron diffusion region, as thick as the 
electron skin depth $\sim d_e$) \cite{pritchett2009_pop}. 
These thicknesses are modified in presence of guide field but the two-scale structure still holds \cite{ricci2004}.

This issue may be resolved in multiple ways. Advanced PIC methods, such as the recently developed Multi-Level Multi-Domain (MLMD) method \cite{innocenti2013,innocenti2014,beck2014}, may be used to locally resolve the electron scales,
while the bigger part of the domain is solved at the ion scales with limited costs. The latter is the approach we follow in our analysis of specific electron dynamics in asymmetric magnetic reconnection.

Direct observations of the electron diffusion region at the magnetopause have been reported in
\textcite{mozer2002} and \textcite{tang2013}, where the presence of a relevant parallel electric fields have been
also observed. 

All these studies have focused on a single isolated reconnection site.
The present work focuses, instead, on the dynamics of the electrons during rapid islands coalescence, in presence of multiple reconnection sites. Two areas are of special interest: 
the islands themselves and the reconnection regions between them. Among the latter, it is of critical importance to distinguish three distinct categories, presenting
the occurrence of vastly different processes. The first type is observed to have the properties of
the asymmetric reconnection mentioned above, similar to the case of a single isolated reconnection site. The second type shows reconnection to occur in vertical (i.e. rotated by 90 degrees with respect to the 
current sheet where the islands develop) conditions between the internal closed field lines of two merging islands. Finally the third 
type show the conditions of two diverging islands with no ongoing reconnection.

Formation of reconnection sites is directly linked with the concept of magnetic islands. In 2D and 2.5D (i.e. two spatial dimensions but three vector components) approaches, 
magnetic islands are formed by the union of two neighboring reconnection outflows.
In 3D, in contrast, the definition of magnetic islands is typically extended as magnetic flux ropes. 
Necessary conditions for magnetic islands to form is the presence of multiple reconnection points, which are in turn the result of multiple local collapses of the current
sheet due to the tearing instability. 
Magnetic islands are moreover thought to be an important source of acceleration for the particles eventually trapped within them \cite{oka2010, drake2006}. 
Given the profound difference in the reconnection outflows evolution 
between the symmetric and 
asymmetric case, magnetic islands in the latter will accordingly show different features in growing with respect to the former, such as the asymmetric outflow swelling and the possible drift of the X-line towards 
the side with the most favorable conditions. So far, most of the studies of the electron dynamics in magnetic islands using simulations have been performed by considering either 
ad-hoc initial profiles including a uniform and regular series of multiple islands (e.g. \textcite{pritchett2007, pritchett2008b}) or pure symmetric conditions (e.g. \textcite{drake2006}, \textcite{oka2010}), while the structure of the islands 
in asymmetric 
case has been recently studied by \textcite{huang2014}. The work presented here instead takes a different approach by investigating the electron dynamics during violent and largely irregular islands merging 
under extreme 
asymmetric conditions, 
following the entire evolution starting from initial formation of multiple reconnection sites.

In order to study this process, two different initial conditions are considered in two different initial current sheets, positioned one above the other at a sufficient
distance to avoid cross talk. The upper current sheet considers the hyperbolic profiles normally 
used in the literature (e.g. \textcite{quest1981}, \textcite{pritchett2008} ), while the lower current sheet is meant to analyze the same asymmetric conditions in presence of a very steep gradient. 
Given its more rapid evolution, the lower current sheet allows us to investigate 
the physical conditions of island merging while comparing with more conventional conditions observed in the upper layer.

The results shown here uncover new important features of reconnection at the magnetopause. The results are especially timely since the recently launched
NASA Magnetospheric Multiscale Mission (MMS) will soon begin its activity by exploring first the magnetopause, looking especially for the regions of electrons dissipation discussed in the
present paper. MMS has unprecedented resolution at the electron scales suggesting to focus our research especially on electron features.

The paper is structured as follows. Section \ref{sec:simulation} describes the initial conditions used in the simulation. Section \ref{sec:overview} gives an 
overview of the
overall results with specific focus on the principal properties of the asymmetric reconnection and the explanation of the approach used in the subsequent sections 
\ref{sec:separatrices} and \ref{sec:islands}. The former describes the 
conditions in the reconnection 
regions and the separatrices, whereas the latter gives more details concerning the situation within coalescing islands. 
Finally, conclusions will be summarized in section \ref{sec:conclusions}.

\section{\label{sec:simulation} Initial Conditions and Simulation Features}

All simulations are performed using the fully kinetic massively
parallel implicit Particle-in-Cell code iPIC3D \cite{markidis2010}, which
allows us to simulate large domains at a
reasonable computational time. In fact, simulations done with the Implicit Moment \cite{vu1992} are 
not required to satisfy the stability constraints of explicit PIC codes \cite{birdsall2004}. Additionally, iPIC3D is extremely scalable up to several thousand of cores \cite{lapenta2012}.

\begin{figure}
 \begin{center}
  \includegraphics[scale=0.23]{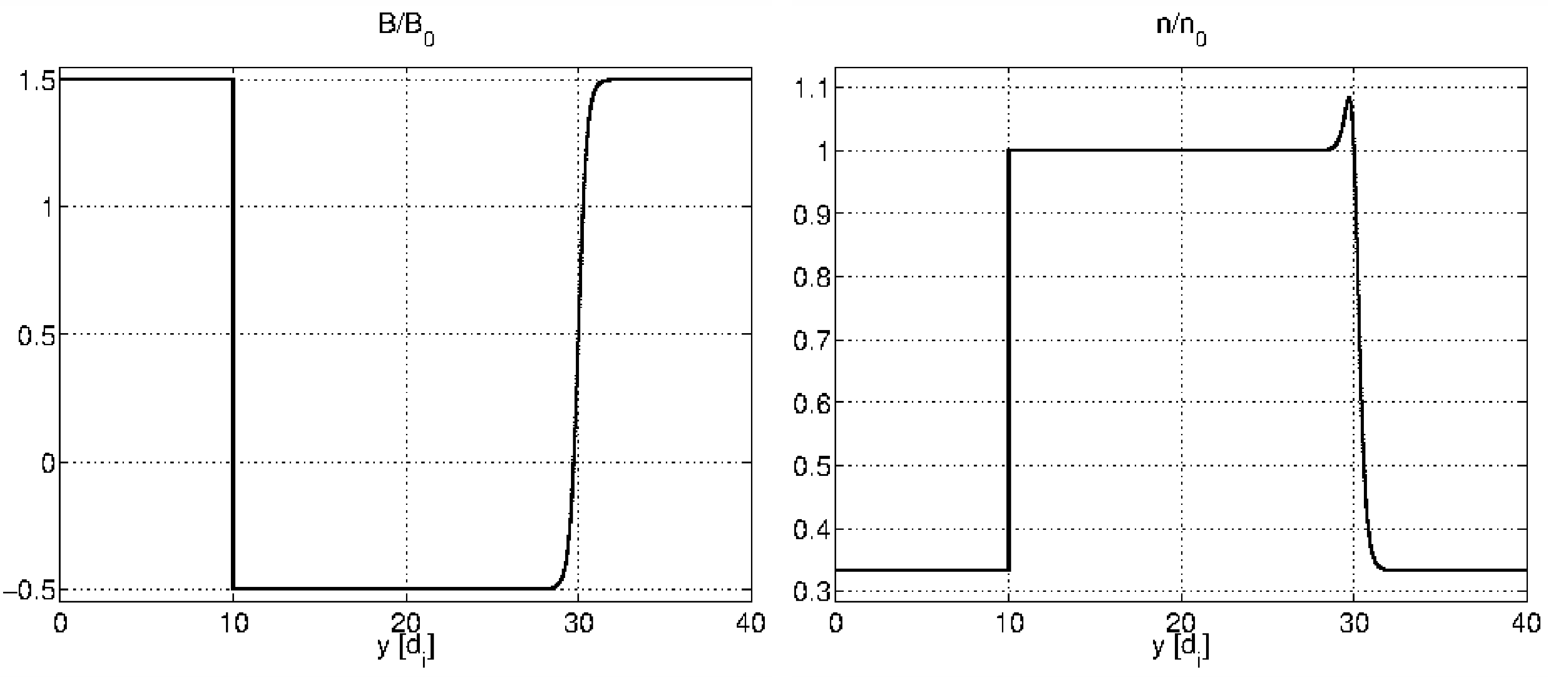}
    \caption{Initial profiles of the magnetic field and density considered in the simulation. The layer at $y_2 \sim 30 \text{ } \unit{d_i}$ simulates the conditions described by the continuous hyperbolic 
  equation \ref{eq:Beq} and \ref{eq:neq}, while the layer set at $y_1 \sim 10 \text{ } \unit{d_i}$ simulates the same asymmetric conditions under a tangential discontinuity.}\label{fig:init_profiles_nB_shifted}
 \end{center}
\end{figure}

We consider the initial profiles for the magnetic field and density given in figure \ref{fig:init_profiles_nB_shifted}, where the two different layers are simulated 
at $y_1 = 10 \text{ and } y_2 = 30 \text{ } \unit{d_i}$. These profiles equations read

\begin{subequations} \label{eq:Beq}
\begin{empheq}[left={ B_{x} \left( y \right) = }\empheqlbrace]{align}
  & \frac{B_0}{2} & \text{$y \le \frac{L_y}{4}$} \\ 
  &  B_0 \left[ \tanh \left(\frac{y - y_2}{\lambda} \right) + R \right]  & \text{$y > \frac{L_y}{4}$} \label{eq:Beqb}
\end{empheq}
\end{subequations}

 \begin{subequations} \label{eq:neq}
\begin{empheq}[left={n\left( y \right) = }\empheqlbrace]{align}
 & n_0 & \text{$y \le \frac{L_y}{4}$} \\
\begin{split} & n_0 \left[ 1  - \alpha \tanh \left(\frac{y - y_2}{\lambda} \right) \right. \\
      &- \left. \alpha \tanh^2 \left( \frac{y - y_2}{\lambda} \right) \right] \end{split} & \text{$y > \frac{L_y}{4}$} \label{eq:neqb}
\end{empheq}
\end{subequations}
where $x$, $y$ and $z$ are, respectively, the North-South, Earth-Sun and
East-West (Dawn-Dusk) direction, and $y_2 = 0.75L_y$.

The upper layer (i.e. at $y = 30 \text{ } \unit{d_i}$ and described by equations \ref{eq:Beqb} and \ref{eq:neqb}) consists of a typical balanced asymmetric current sheet often reported
in the literature (see e.g. in \textcite{quest1981}  and  \textcite{pritchett2008}). 

The lower layer is meant to study a very thin current sheet by setting a strong discontinuity (i.e. a step function centered in $y_2 = 0.25 L_y$) in both density and magnetic field. 
This configuration resembles the Riemann's problem considered in the 
earliest theories on magnetopause \cite{ferraro1952} and further MHD models for the study of single tangential discontinuities standing in this region when the magnetic field is northward 
(e.g. \textcite{biernat1989}).

All quantities are normalized to the magnetosheath conditions, including the ion skin depth  $d_i = \frac{c}{\omega_{pi}}$ ($c$ is the speed of light and $\omega_{pi}$ is the ion plasma frequency), 
whose value, given a density of nearly $30\ \unit{cm^{-3}}$ (e.g. \textcite{cassak2007}), corresponds to a physical
distance of $\sim 42\ \unit{km}$. The current sheet half-thickness is here fixed to $\lambda = 0.5 d_i$.
In order to satisfy the pressure balance, and choosing a ion-electron temperature ratio $ T_i = 2 T_e$, we obtain $R = \frac{1}{2}$ and $\alpha = \nicefrac{1}{3}$. 
To save computational time we follow the literature and use a 
mass ratio $m_r = \nicefrac{m_i}{m_e} = 256$ by rescaling all related quantities. The temperature profile is kept constant along the sheet, as often done in the
literature (e.g. \textcite{swisdak2003}, \textcite{pritchett2008}). These profiles do not give an exact solution of the Vlasov's equation leading
to a imperfect pressure balance, as pointed out by \textcite{pritchett2008}. 

A double periodic  $\left( 40 \times 40 \right) d_i$ box is simulated, with $2048 \times 2048$ cells and $dt \sim 5.5\cdot 10^{-4} \omega_{c,i}^{-1}$, where $ \omega_{c,i}$ is the ion gyrofrequency.
This leads to a spatial resolution $dx \sim 3 \text{ } \unit{d_e}$, where $d_e$ is the electron skin depth.

In order to initialize the reconnection, a perturbation is applied on the upper layer.
We considered as perturbation the equation used in \textcite{lapenta2010}, namely

\begin{equation} \label{eq:eq_pert}
\begin{aligned}
\delta A = A_{z0} & \cos \left( \frac{2 \pi x}{L_{\Delta}} \right) \cos \left( \frac{2 \pi \left( y - y_2 \right)}{L_{\Delta}} \right) \\
& \times \exp \left( \frac{- \left(x^2 + \left( y - y_2 \right)^2 \right)}{\sigma^2} \right) 
\end{aligned}
\end{equation}
where $\sigma$ gives the strength of the perturbation, $L_{\Delta} = 10 \sigma$ and $y_2 = 0.75 L_y$. A value of $\sigma = \nicefrac{d_i}{2}$, i.e. as thick as the current sheet, 
is sufficient to trigger the reconnection without influencing the lower layer.
No perturbation is set on the lower layer instead, given the strong tearing
instability self-generated by the profiles.

\begin{figure}
 \begin{center}
  \includegraphics[scale=0.35]{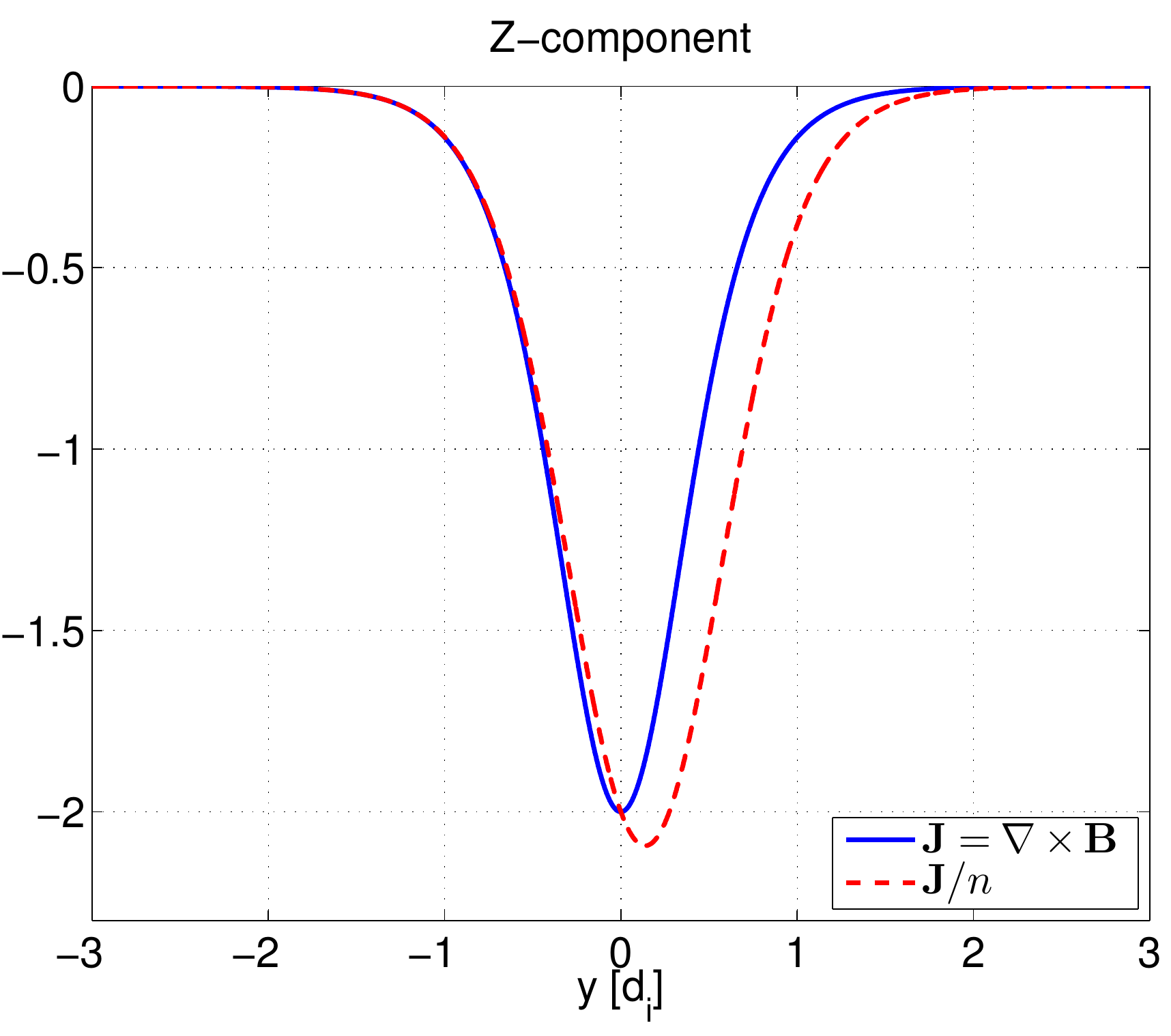}
    \caption{Out-of-plane component of the current (blue continuous line) and of the drift velocity (red dotted line). The ion drift has been neglected and the current is totally carried by the electrons.}\label{fig:drift_vel}

 \end{center}

\end{figure}

Unlike the symmetric configuration where the fluid drift velocity is constant
over the domain, in this case a variation in the drift velocity is established due to the density asymmetry.
According to the Ampere's law, an out-of-plane current density must be included, as 

\begin{equation} \label{ampere_law}
 \mathbf{J} = \frac{\nabla \times \mathbf{B}}{\mu_0} = - \frac{B_0}{\lambda}
 \sech^2 \left( \frac{y}{\lambda} \right) \cdot \mathbf{\hat{e}_z}
\end{equation}
Considering the asymmetric density, the shifted
velocity profile shown in figure \ref{fig:drift_vel} is thus obtained.

Following previous studies \cite{pritchett2008}, ion drift has been neglected and all the current is carried by electrons.

The current at the lower layer is initially set to zero, making this layer intrinsically unstable, as suggested in previous literature (e.g. \textcite{markidis2013}).
On the other hand, this only represents the initial configuration, 
as a strong current layer rapidly grows across the discontinuity at the beginning of the simulation.

\section{\label{sec:overview} Overview of the Results}

This section gives an overview of the main results obtained from the global simulation.

\begin{figure}
 \begin{center}
  \includegraphics[scale=0.8]{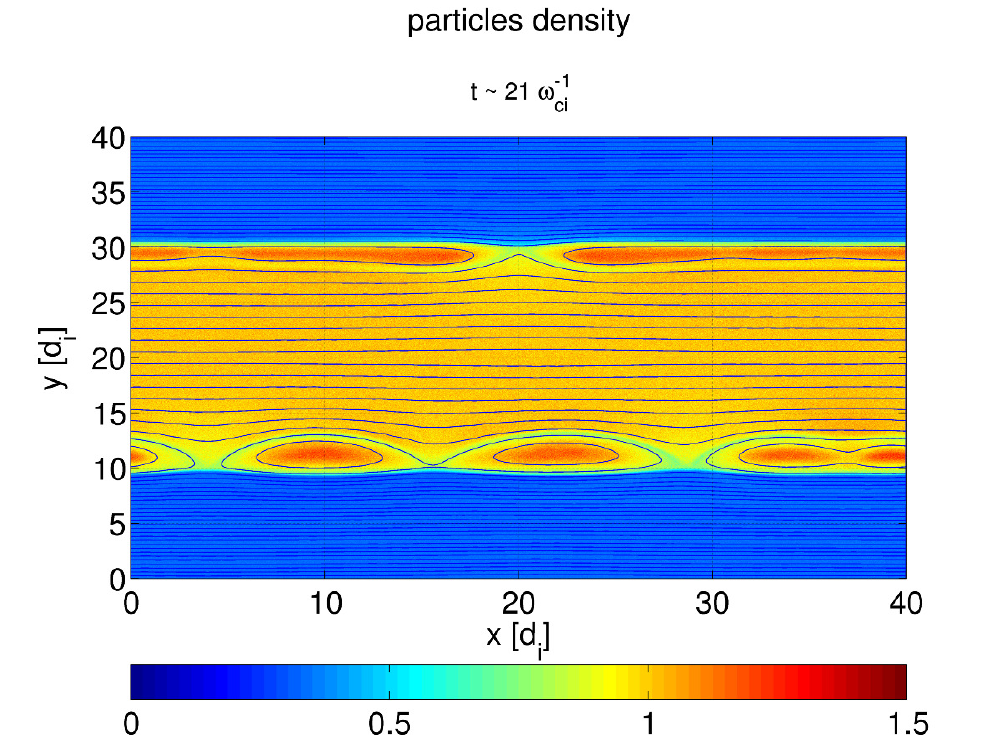} 
  \caption{Electrons density at $\sim 21 \text{ } \unit{\omega_{c,i}^{-1}}$ in the two different layers. An advanced stages of the island merging is visible in the lower layer, whereas 
  in the upper layer the reconnection is at its onset.} \label{fig:rho_cycle38000pic}
 \end{center} 
\end{figure}

Figure \ref{fig:rho_cycle38000pic} shows the particles density at $t \sim 21\ \unit{\omega_{c,i}^{-1}}$ in order to highlight the differences between the two layers. 
The top layer shows some of the well-known signatures of asymmetric reconnection. Even though still at its onset, one can clearly notice the predominant swelling of the plasmoids 
toward the weakest magnetic field region \cite{cassak2007,pritchett2008,huang2014}.

\begin{figure}
 \begin{center}
  \includegraphics[scale=0.8]{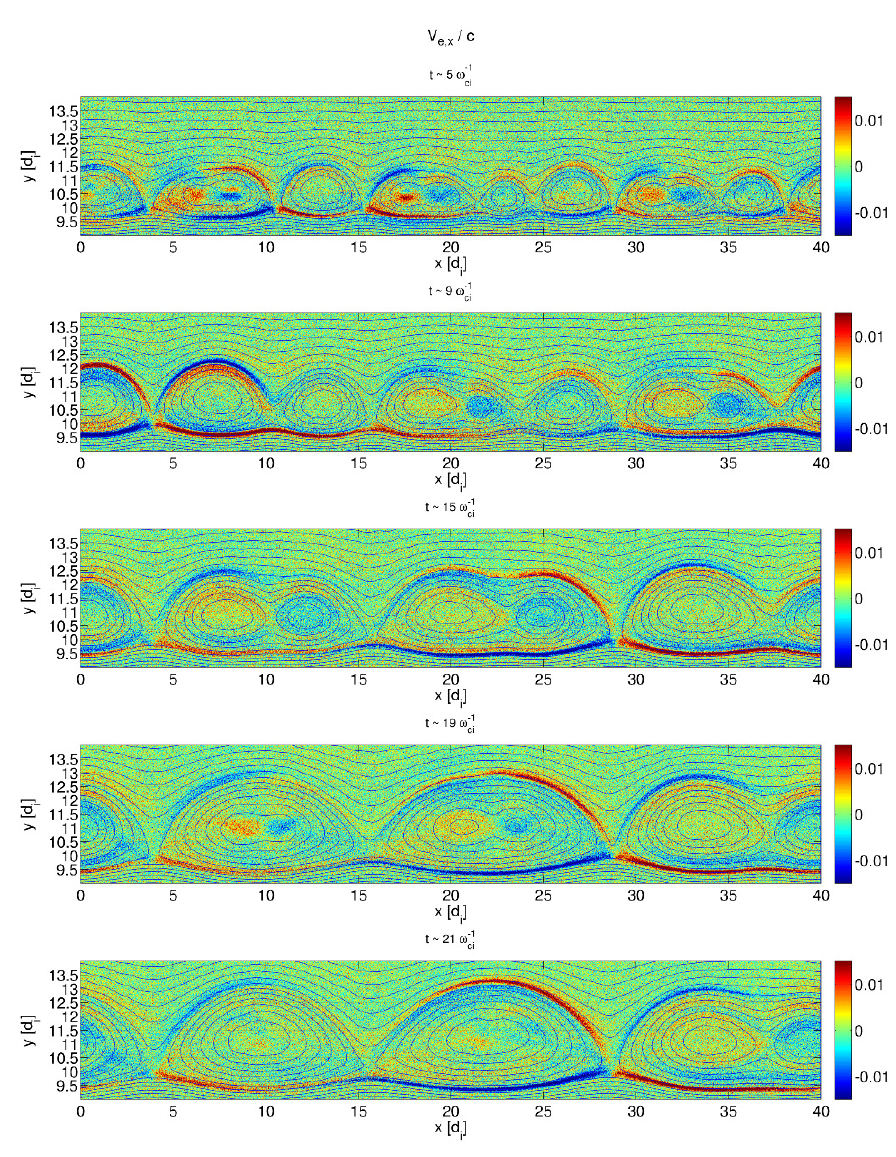} 
    \caption{$x$ component of the electron velocity (normalized to the speed light) at different times, respectively $\sim 5 \text{ } \unit{\omega_{c,i}^{-1}}$, $\sim 9 \text{ } \unit{\omega_{c,i}^{-1}}$, $\sim 15 \text{ } 
    \unit{\omega_{c,i}^{-1}}$, 
    $\sim 19 \text{ } \unit{\omega_{c,i}^{-1}}$ and $\sim 21 \text{ } \unit{\omega_{c,i}^{-1}}$. Notice the rapid islands growth and coalescence.}
    \label{fig:B_zoom}
 \end{center}
\end{figure}

The lower layer is examined further in figure \ref{fig:B_zoom} where the $x$ component of the electron velocity is plotted at different times ($\sim 5 \text{ } \unit{\omega_{c,i}^{-1}}$, $\sim 9 \text{ } \unit{\omega_{c,i}^{-1}}$, $\sim 15 \text{ } 
    \unit{\omega_{c,i}^{-1}}$, 
    $\sim 19 \text{ } \unit{\omega_{c,i}^{-1}}$ and $\sim 21 \text{ } \unit{\omega_{c,i}^{-1}}$).

Due to the considerable gradients and the presence of a strong non-equilibrium condition, as well as the extreme thinness of the sheet, violent and multiple reconnection events
soon take place in this layer with consequent rapid island merging. The growing magnetic islands visible  at later stages in 
figure \ref{fig:B_zoom} are then the result of a consecutive merging of the reconnection outflows, followed by the subsequent coalescence between the 
resulting islands.
The formation of multiple reconnection points is explained with the collapse of the layer caused by the tearing instability \cite{drake1977}.
Some of these initial reconnection points eventually become  weak and disappear, generating the mutual attraction of the nearest islands which ultimately lead to the merging \cite{oka2010}. 
In coalescing together, two magnetic islands can generate  secondary vertical reconnections \cite{pritchett2008b}, sometimes called in the literature: anti-reconnections
\cite{oka2010}. The signature of multiple reconnection points and the presence of coalesced magnetic islands have also been  observed in the magnetotail (e.g. \textcite{eriksson2014,eriksson2015},
\textcite{eastwood2007}), as well as the equivalent 3D version known as magnetic flux ropes
 at the dayside magnetopause (e.g. \textcite{hasegawa2010,oieroset2011,oieroset2014}).

While the situation just described has not been previously considered in asymmetric cases,  symmetric \cite{lin1995}
and mono-dimensional asymmetric conditions \cite{omidi1995} have been reported previously. Island merging  and the resulting effect on electron acceleration in symmetric conditions has been studied in
previous works (\textcite{pritchett2007,pritchett2008b},\textcite{wan2008},\textcite{oka2010}).

Given the violent nature of the phenomena in the lower layer, the focus of the rest of the study will be on  the electron dynamics in this region.

Particles gain energy from two main processes while being trapped within a single magnetic island. 
This first mechanism has been investigated in \textcite{drake2006} and \textcite{huang2014}. In particular, \textcite{drake2006} pointed out that
the Fermi acceleration
mechanism is responsible for the increase of the particle parallel energy during  island contractions. The internal structure of a
magnetic islands in both the symmetric and asymmetric conditions has been studied in \textcite{huang2014}. They 
demonstrated that in asymmetric configurations the internal current is predominantly carried by the electrons running along the separatrices from the side with the highest density towards 
 the lowest. 
This internal pattern leads to the second process of acceleration. In flowing along the separatrices, the electrons 
experience an acceleration when they run in the proximity of the two lateral X-lines of the island, where the strong out-of-plane electric field formed earlier due to the island contraction is present. 
Further efficient particles acceleration is thought to occur after the latest stages of the coalescence, when the islands become particularly large
\cite{pritchett2008b}.

\begin{figure}
 \begin{center}
  \includegraphics[scale=0.8]{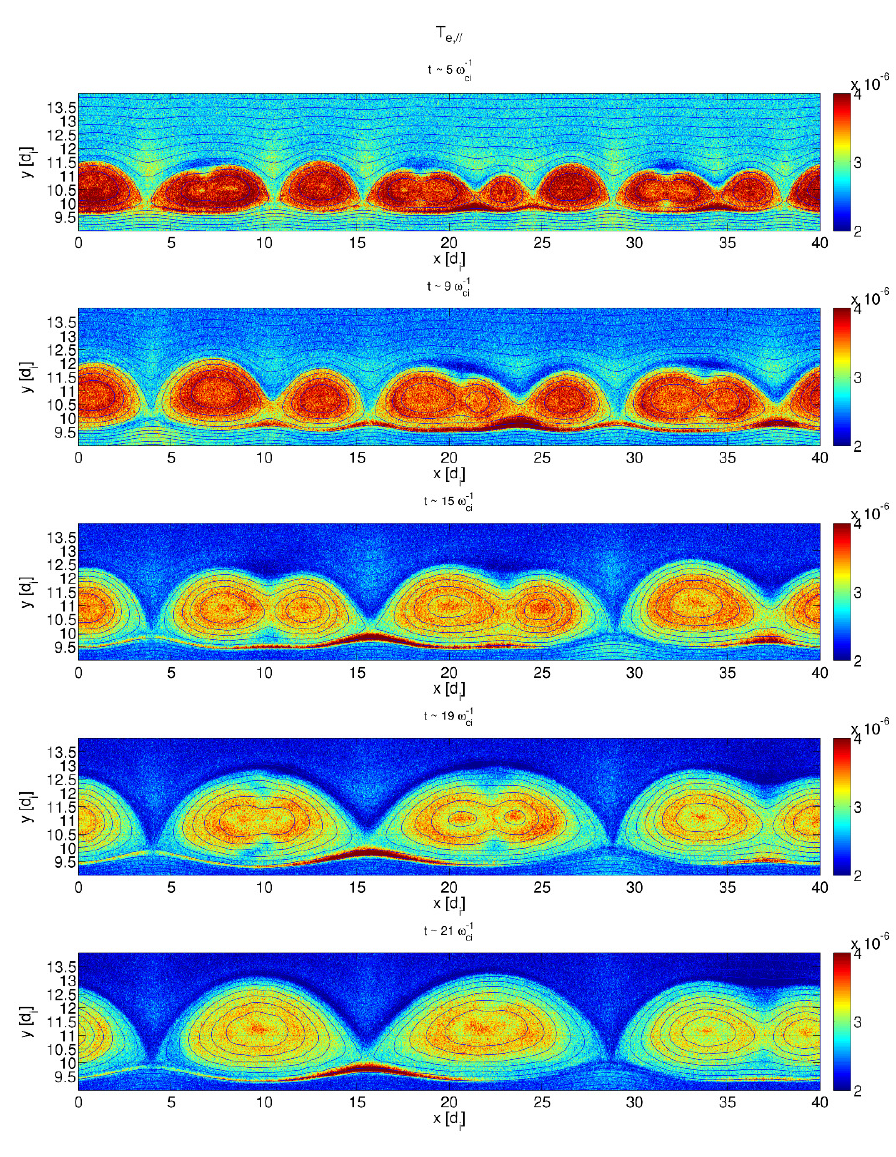} 
    \caption{Parallel component of the electrons temperature $T_{e_{\|}}$ at different times. These plots underline the large heating in some specific regions
    of the lower side, such as at $x \sim 17 \text{ } \unit{d_i}$ (Multimedia View).} \label{fig:T0par}
 \end{center}
\end{figure}

\begin{figure}
 \begin{center}
  \includegraphics[scale=0.8]{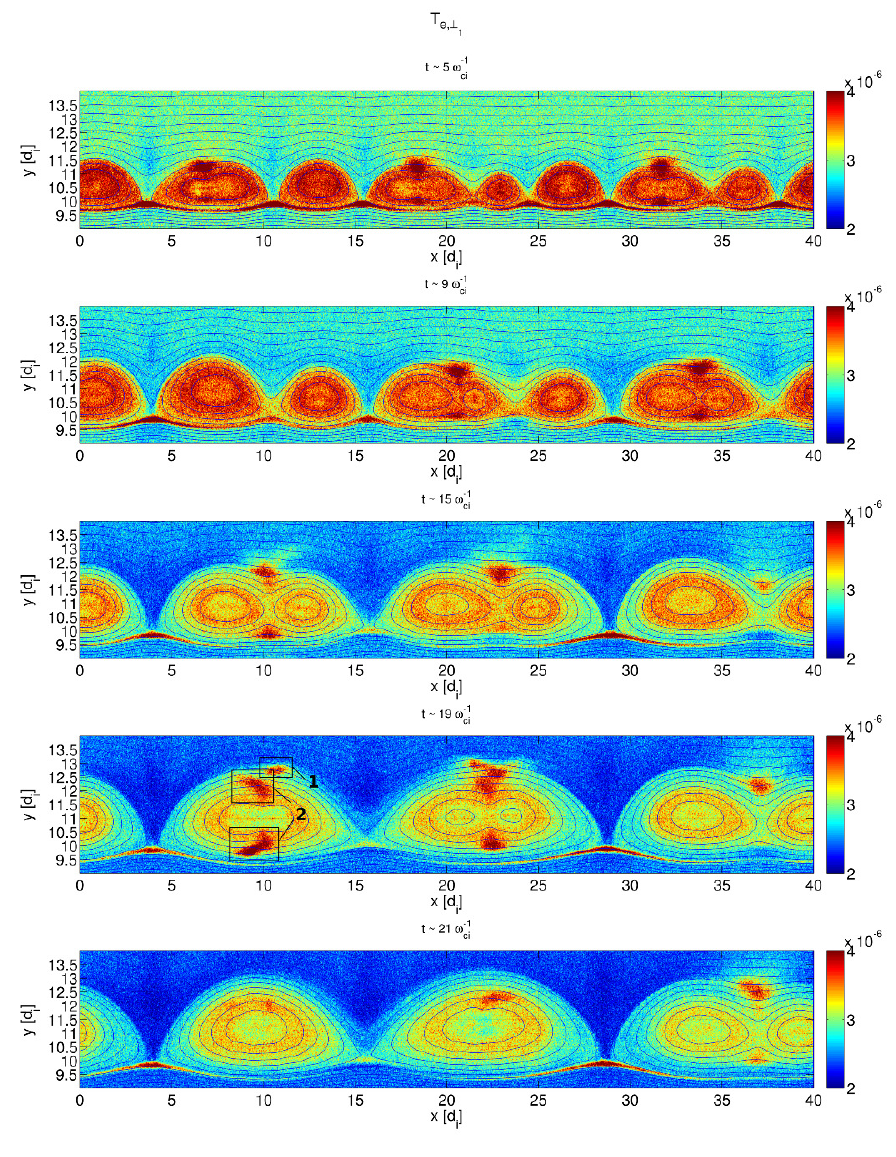} 
    \caption{In-plane perpendicular component of the electron temperature $T_{e_{\perp 1}}$ at different times. 
    These plots show some features not observed in the parallel component, such as those patches marked with the black box $1$ and $2$. Further analysis have proved that these two patches are 
    actually moving in the opposite directions (Multimedia View).} \label{fig:T0perp1}
 \end{center}
\end{figure}

\begin{figure}
 \begin{center}
  \includegraphics[scale=0.8]{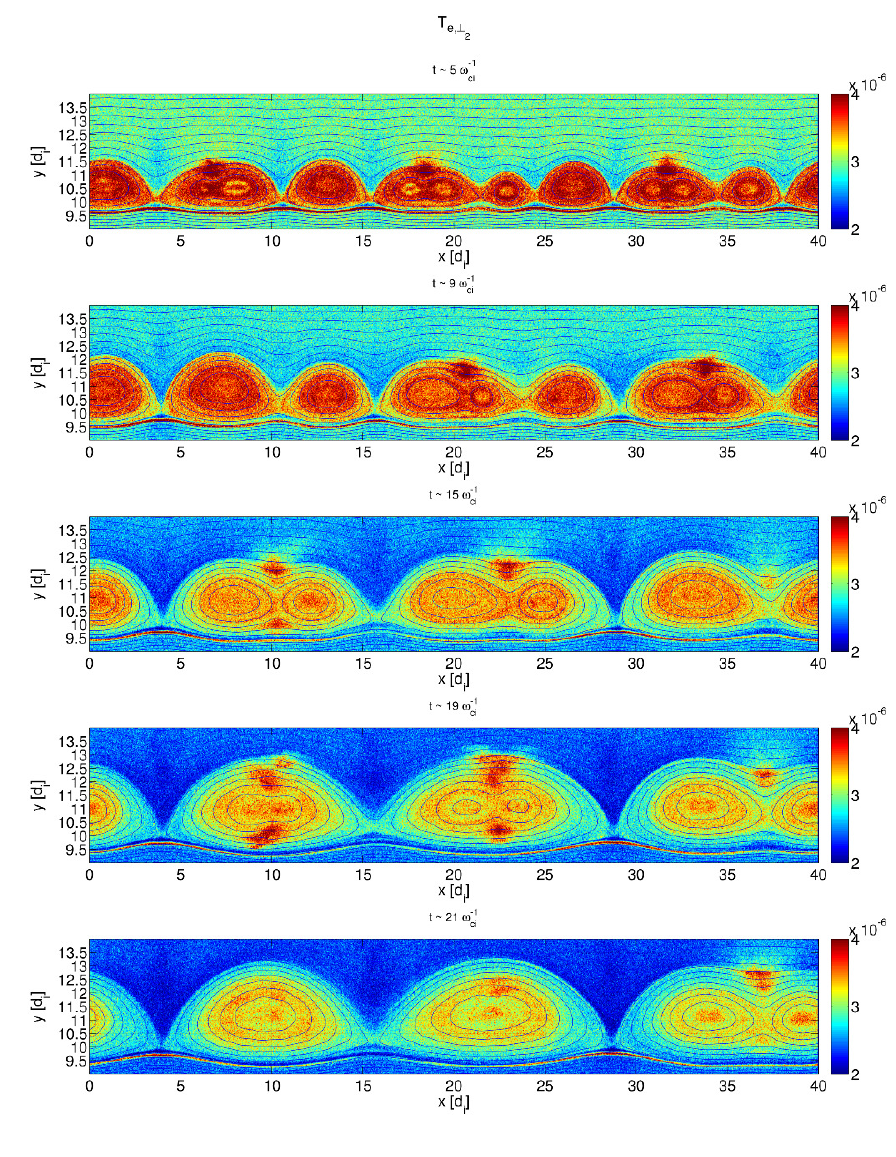} 
    \caption{Out-of-plane perpendicular component of the electrons temperature $T_{e_{\perp 2}}$ at different times. 
    These plots further remark the presence of the patches mentioned for $T_{e_{\perp 1}}$. Additionally, this component shows an evident heated layer all over the lower side (Multimedia View).} \label{fig:T0perp2}
 \end{center}
\end{figure}

We compare the parallel and perpendicular electron temperatures at different time steps 
in figures \ref{fig:T0par}, \ref{fig:T0perp1} and \ref{fig:T0perp2} (Multimedia View). This new coordinate system is based on the local magnetic field lines direction.
In particular, the parallel quantity considers the component aligned to the local direction of the field direction. The in-plane $\perp_{1}$ component is aligned to the $\mathbf{B} \times \mathbf{z}$ 
direction still lying in the $x-y$ plane, whereas $\perp_{2}$  is the third orthogonal out-of-plane component.

We observe the two perpendicular components of the temperature to be dominant  at the separatrices bordering with the side of lower density and higher magnetic field, mostly at the 
earliest stages (e.g. at $\sim 5 \text{ and } \sim 9 \text{ } \unit{\omega_{c,i}^{-1}}$). However, $T_{e_{\perp_{1}}}$ is later observed to remain localized after the coalescence of larger islands, while instead
$T_{e_{\perp_{2}}}$  shows a (quasi)-uniformly heated layer connecting all the  lower separatrices.

In contrast, the parallel component shows from the beginning a dominant heating only in the regions where reconnection is observed to occur.
Additionally, we notice the formation of an alternate pattern between the parallel and in-plane perpendicular components of the electron temperature in the proximity of the  X-lines. 
The latter are clear in the two central regions at $x \sim 15 \text{ and } \sim 28 \text{} \unit{d_i}$ of the last three panels in figures \ref{fig:T0par}, \ref{fig:T0perp1} and \ref{fig:T0perp2},
where an alternate dominance between the parallel (i.e. $x \sim 15 \text{ } \unit{d_i}$) and the perpendicular components (i.e. $ \sim 28 \text{ } \unit{d_i}$) is glaring.

Finally, particularly interesting are the localized heating patches visible within some islands, mostly in the perpendicular components,
and identified by a black box in figure \ref{fig:T0perp1}. 
This peculiarity will be analyzed in section \ref{sec:islands} and will be shown to occur at the meeting point of two merging islands. 
The heating can be a clear signature of what has been earlier mentioned as 
anti-reconnection.

\section{Reconnection Regions and Separatrices} \label{sec:separatrices}

\begin{figure}
 \begin{center}
  \includegraphics[scale=0.8]{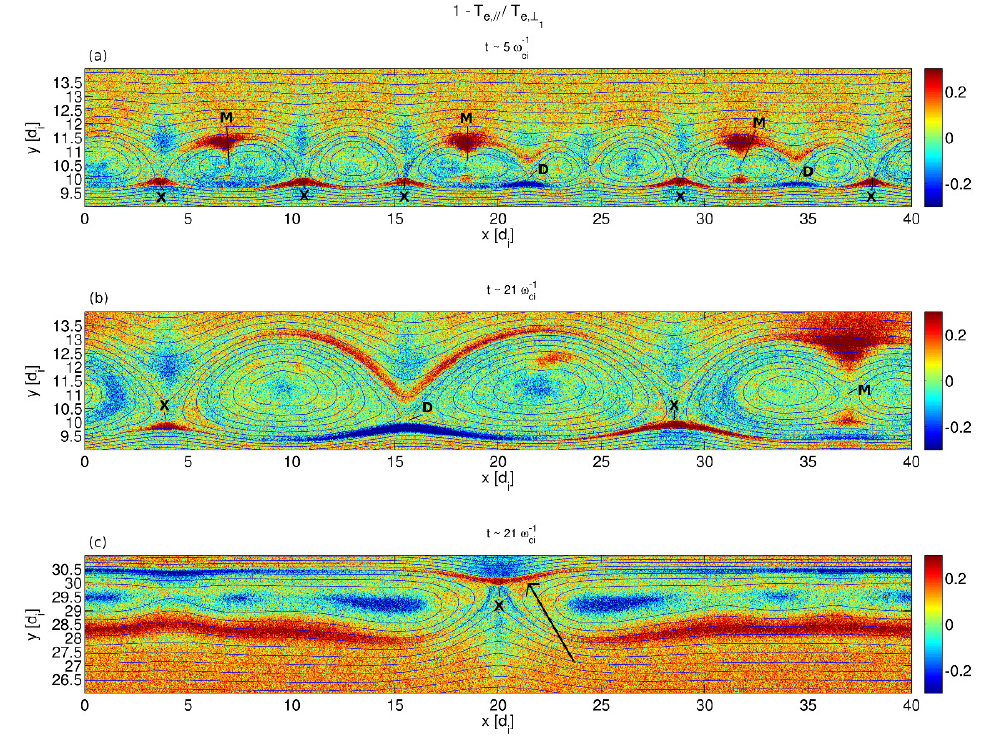} 
    \caption{Plot of the quantity $1 - \frac{T_{e,\|}}{T_{e,\perp_{1}}}$ to analyze the presence of anisotropies in the system. Panel a and b show the situation in the lower layer at, respectively, 
    $t \sim 5 \text{ } \unit{\omega_{ci}^{-1}}$ and $t \sim 21 \text{ } \unit{\omega_{ci}^{-1}}$. Panel c represents the upper layer at $t \sim 21 \text{ } \unit{\omega_{ci}}$ for a direct comparison.
    Three different type of regions are distinguished and marked with the letters X, M and D.} \label{fig:anisotropyTOT}
 \end{center}
\end{figure}

\begin{figure}
 \begin{center}
  \includegraphics[scale=0.8]{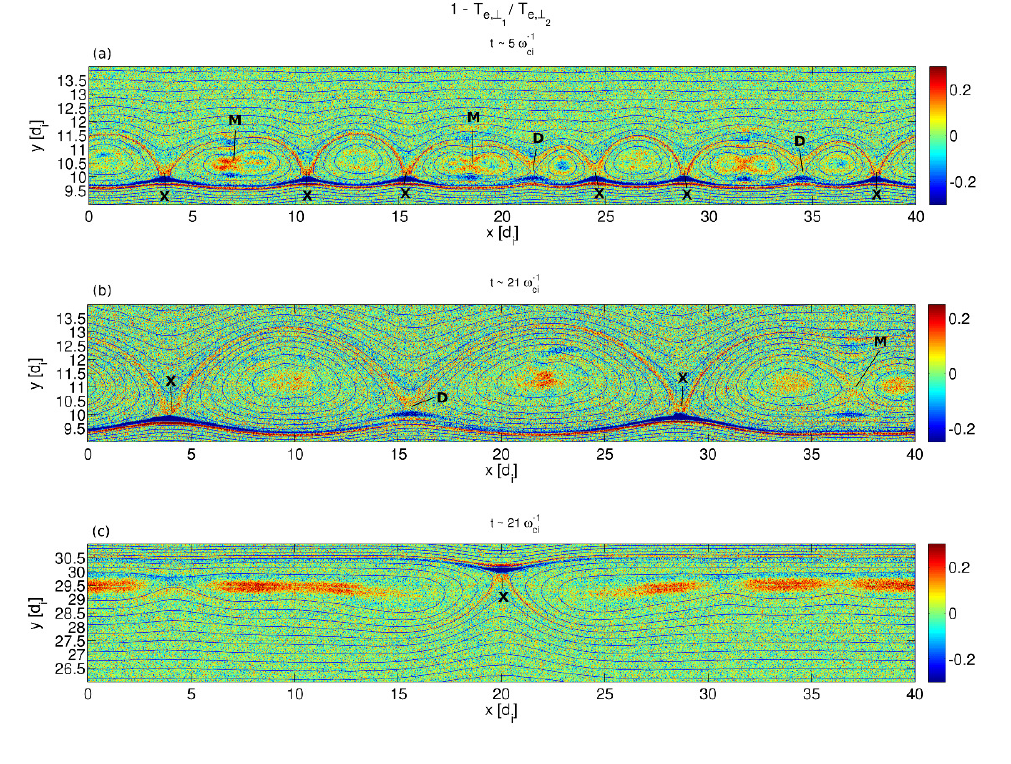} 
    \caption{Plot of the quantity $1 - \frac{T_{e,\perp_{1}}}{T_{e,\perp_{2}}}$ to analyze the presence of non-gyrotropies in the system at the same times as in figure 
    \ref{fig:anisotropyTOT}.} \label{fig:gyrotropyTOT}
 \end{center}
\end{figure}

In order to study the different behavior of electrons in the reconnection regions and in the separatrices, in figure 
\ref{fig:anisotropyTOT} and \ref{fig:gyrotropyTOT}
the quantity $1 - \nicefrac{T_{\|}}{T_{\perp_{1}}}$ and  $1 - \nicefrac{T_{\perp_{1}}}{T_{\perp_{2}}}$ are plotted. Three panels are shown: panels a and b represent the lower layer at two different times 
(i.e. $t \sim 5 \text{ } \unit{\omega_{ci}^{-1}}$ and $t \sim 21 \text{ } \unit{\omega_{ci}^{-1}}$), 
while panel c allows us to directly compare the lower layer with the well known situation 
of the asymmetric reconnection in the upper layer at $t \sim 21 \text{ } \unit{\omega_{ci}^{-1}}$. 

From these plots
we can distinguish three main regions of interest, namely those marked with X, M and D in the figure.
Thanks to the comparison with the situation in the upper layer (panel c), we can deduce that regions marked with X define the asymmetric reconnection events occurring in the horizontal direction with the clear vertical particles inflow from the top and the bottom. 
Regions marked with M instead define those reconnection regions occurring between two merging islands, which also include those events often identified as 
anti-reconnections \cite{pritchett2008,oka2010}. We preferred avoiding this latter nomenclature as they do not show remarkable physical differences with respect to those 
in the region X, except for the fact they occur rotated in space with respect to the horizontal inflow. 
Finally, the third type of regions are marked with D. 
These regions show a similar pattern to regions X but with a few peculiar  features. In particular, we distinguish this third region from the others based on the strong opposite anisotropy observed with respect
to regions X.  
Also, we notice a secondary double agyrotropic structure forming in these regions exactly above that already existing (see e.g. around the D-type at $x \sim 16 \text{ } \unit{d_i}$ in panel b in 
figure \ref{fig:gyrotropyTOT}).

\subsection{X-type reconnection sites}

Thanks to the direct comparison between the two different current sheets shown in panels b and c in figures \ref{fig:anisotropyTOT} and \ref{fig:gyrotropyTOT},
we are able to establish that the types X represent the typical reconnection in asymmetric conditions. 
The situation is characterized by the  presence of an anisotropic particle inflow. This anisotropy is present also  in the symmetric reconnection case, but here the top-bottom symmetry is of course broken. 

The inflow anisotropy is explained by the conservation of the magnetic moment as a consequence of the combined decrease of the in-plane perpendicular magnetic field component and of the overall magnetic field strength 
due to the ongoing reconnection. Electrons in this area are in fact still magnetized. 
This mechanism is commonly known as adiabatic heating, and such effect in this region has been explained in \textcite{egedal2009,egedal2011}. 
A significant reduction of $T_{\perp_{1}}$ in the upper inflows  
is also visible in figure \ref{fig:T0perp1} (see e.g. inflows in panel a). Likewise, a low-$T_{\perp_{1}}$ inflows are observed in the side with stronger magnetic field, 
which show to be strongly squeezed due to 
the different resistance
encountered by the plasma on this side. 
An especially interesting feature is the strong anisotropic narrow layer standing between the squeezed inflow and the reconnection region, appearing only on the magnetospheric side, 
where $B$ is greater and $n$ lower (pointed with an arrow in figure \ref{fig:anisotropyTOT}, panel c). 
A similar finding has been observed and studied in in \textcite{egedal2011}.


\subsection{M-type reconnection sites}

Reconnection sites of the type M are observed to occur between closed magnetic field lines of merging islands. Given the strong 
non-uniformity of this current sheet, these reconnection events can either occur in symmetric or asymmetric conditions depending on the history of previous mergers of   islands at earlier times. This explains the fact that in some regions the particles inflows, which is now horizontal, is symmetrically anisotropic, while in others 
they asymmetrically anisotropic, as seen above in regions X. However, some peculiarities seen in regions X, such as the strong anisotropic layer bordering the side with the higher magnetic field, are not observed in  
regions of type M.

\begin{figure}
 \begin{center}
  \includegraphics[scale=0.9]{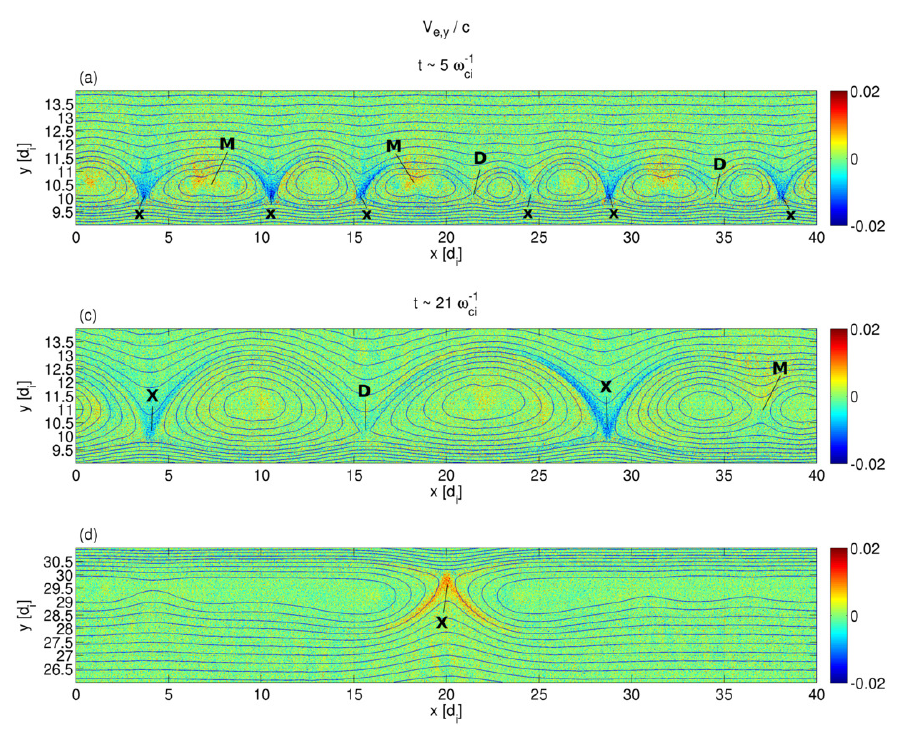} 
    \caption{$y$-component of the electron velocity in the two layers and at two different times, showing a dominance of the motion towards from the X-line from the side with the lower magnetic field.} \label{fig:V0y}
 \end{center}
\end{figure}

The distinguishing  hallmark of the M regions are the anisotropic asymmetric vertical outflows observed to leave 
the merging points (figure \ref{fig:anisotropyTOT}). For clarity, the asymmetry just mentioned concerns  the difference in the both heating and outflows with respect to the horizontal $y-z$ 
plane adopted in this
simulation, and not the asymmetries of the initial profiles studied in this work.
The $y$-component of the electron velocity in figure \ref{fig:V0y} shows that the outflows are in fact preferentially released upwards where the magnetic field is weaker. 
However, smaller jets are also detectable downward, 
although they seem to be confined and inhibited by the underlying layer. 

The anisotropy of these jets is confirmed by the strong perpendicular components clearly visible in figure \ref{fig:T0perp1} and \ref{fig:T0perp2}. 
Such considerable perpendicular heating in the outflows resembles that observed in the symmetric case at the dipolarization fronts
\cite{lapenta2011,lapenta2014}. 
Dipolarization fronts are mainly caused by the strong interaction between the hot plasma and magnetic fields flowing out from the reconnection point and
the unperturbed ambient plasma. Due to the presence of an internal localized non-uniform density, interchange instabilities can be excited ultimately leading to the 
generation of hot particles \cite{lapenta2011b}. 
However, a direct correlation between the outflow velocity (figure \ref{fig:V0y}) and the heating anisotropy is only confined to the very first layers, as $V_y$ decreases and vanishes rapidly in a very
short range.

\subsection{D-type reconnection sites}

The regions marked with D show an opposite behavior with respect to the regions X, together with some additional features.

Unlike regions X, these regions no longer show an increase of the anisotropy in the lower inflow, 
which seems instead to be included in the underlying anisotropic layer. 
A top-bottom symmetry in the anisotropy along the islands profile is evident in the D regions, with a clear dominance of $T_{\|}$
at the bottom and a dominance of $T_{\perp_{1}}$ at the top. With the help of figures \ref{fig:T0par} and \ref{fig:T0perp1}, the latter process is better 
demonstrated by the particular low value of $T_{\|}$, while the former is explained with the dominance of $T_{\|}$ and the small value of  
 $T_{\perp_{1}}$.

\begin{figure}
 \begin{center}
  \includegraphics[scale=0.88]{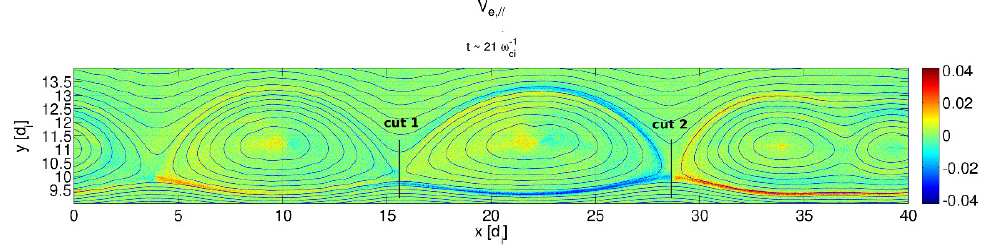} 
    \caption{Electron parallel velocity at $t \sim 21 \text{ } \unit{\omega_{ci}^{-1}}$. This figure shows the flow of the electrons, which is predominantly observed along the 
    separatrices. In particular, electrons are seen to converge toward the X-lines (region X) and reach the D regions from the lower separatrices, where they ultimately interact with each other leading to
    a parallel heating caused by local streaming instabilities. Lines marked with cut 1 and cut 2 indicate the vertical-cuts considered for the phase spaces in figure \ref{fig:recregTOT}.} \label{fig:VeparPerp}
 \end{center}
\end{figure}

In particular, the analysis of the parallel electrons velocity given in figure \ref{fig:VeparPerp} reveals that in these  the islands are diverging.
The simultaneous absence of a vertical inflow (figure \ref{fig:V0y}) and a relevant agyrotropy in the tiny central region (figure \ref{fig:gyrotropyTOT}, e.g. panel b), leads us to believe
that a reconnection event is not occurring in that moment in the D regions.


Nevertheless, we still need to investigate why the parallel component is so strong in the lower part of the D regions. 
Parallel heating in the vicinity of the reconnection sites has been observed in symmetric simulations \cite{egedal2009,egedal2012} and observations 
(e.g. in \textcite{egedal2005} and  from the Cluster satellites in \textcite{egedal2012}).
In classical symmetric conditions the increase of $T_{\|}$ has been explained with the Fermi acceleration occurring within a narrow region of strong parallel electric 
field \cite{egedal2009}. 

\begin{figure}
 \begin{center}
  \includegraphics[scale=0.8]{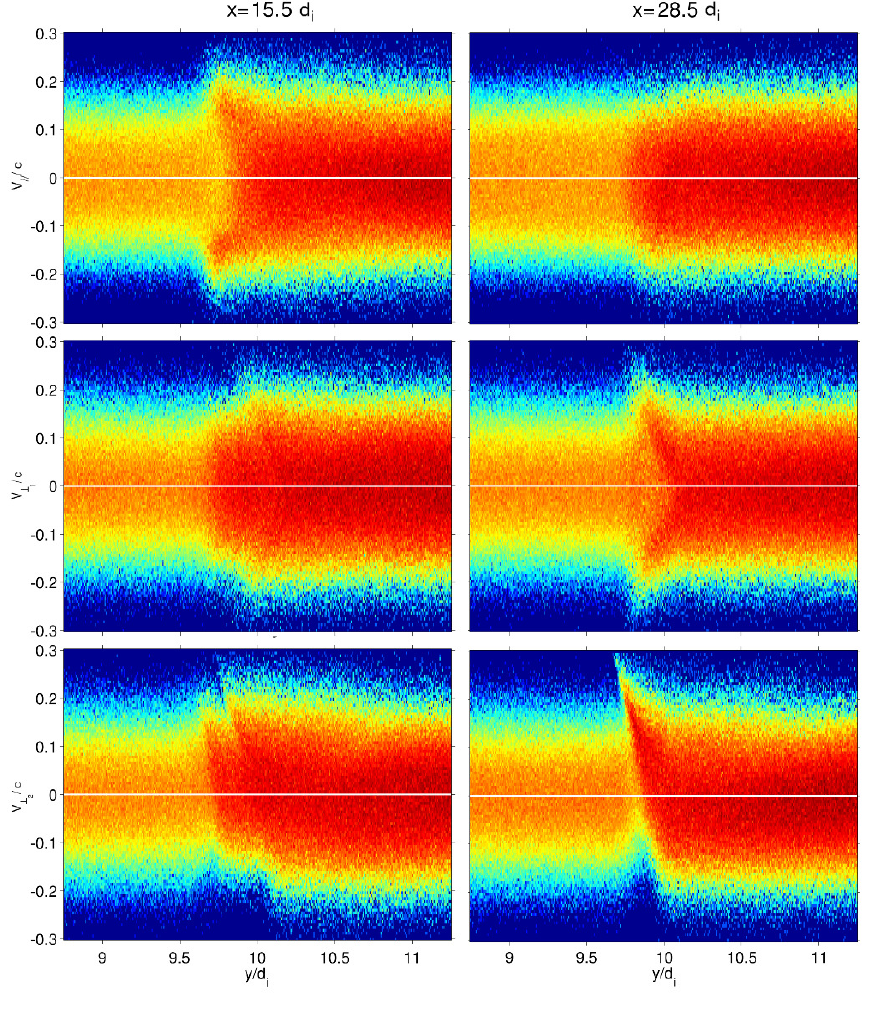} 
    \caption{Comparison of the electrons phase spaces along two vertical-cuts at $x \sim 15.5 \text{ (region D) and } \sim 28.5 \unit{d_i}$ at $t \sim 21 
    \text{ } \unit{\omega_{ci}^{-1}}$ (region X). 
    The different behavior is clearly visible in all the three components. In particular, the ongoing reconnection is further confirmed by the strong out-of-plane perpendicular component in the 
    right column (region X).} \label{fig:recregTOT}
 \end{center}
\end{figure}

A comparison of the electron behavior in regions X and D is shown through 
the phase spaces in figure \ref{fig:recregTOT} obtained from a vertical cut at $x \sim 15.5 \text{ and } \sim 28.5 \text{ } \unit{d_i}$. 
The left column refers to the D regions and indicates that particles experience, within the tiny range $ 9.7 \le y \le 10 \text{ }\unit{d_i}$, a parallel asymmetric acceleration to either
positive or negative values (with maximum at $ y \sim 9.7 \text{ }\unit{d_i}$), a thermal heating in the in-plane perpendicular direction (with the maximum value at $ y \sim 9.9\text{ }\unit{d_i}$) and
a double acceleration in the out-of-plane perpendicular direction (with  acceleration peaks around $y \sim 9.7 \text{ and } y \sim 9.8 \text{ } \unit{d_i}$). 
The
situation is different in the regions X described in the right column. Here particles are observed to experience, within the same narrow range, a weak simultaneous acceleration
and heating in the parallel direction (with maximum at $ y \sim 10\text{ }\unit{d_i}$) and a relevant increment of the thermal energy in the in-plane
perpendicular component  (with maximum at $ y \sim 9.8\text{ }\unit{d_i}$). 
Finally, $V_{\perp_{2}}$ shows the presence of a strong acceleration in the out-of-plane direction, clear sign of an ongoing reconnection with no guide field  (with the extreme 
maximum at $ y \sim 9.75$).

\begin{figure}
 \begin{center}
  \includegraphics[scale=0.22]{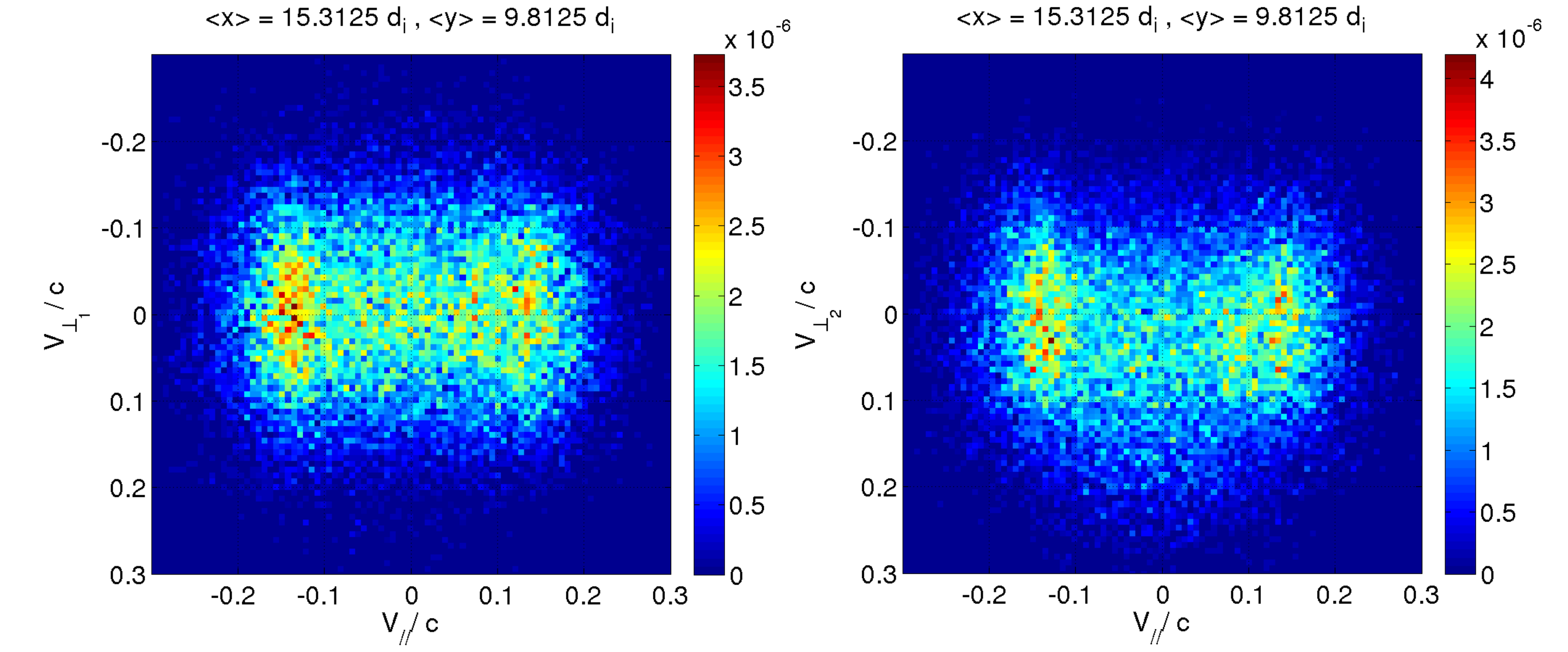} 
    \caption{Electron velocity distributions in the region D at $x \sim 15.5 \text{ and } y \sim 10 \text{ } \unit{d_i}$ at $t \sim 21 \text{ } \unit{\omega_{ci}^{-1}}$ (the average centered values are 
    pointed out on the top of the figures). This region is the main crossroad 
    for the particles coming from the nearest reconnection regions (i.e. regions X) and flowing along the separatrices. Two opposite beams are in fact clearly visible in this region.} \label{fig:VparVperp1recreg}
 \end{center}
\end{figure}

\begin{figure}
 \begin{center}
  \includegraphics[scale=0.8]{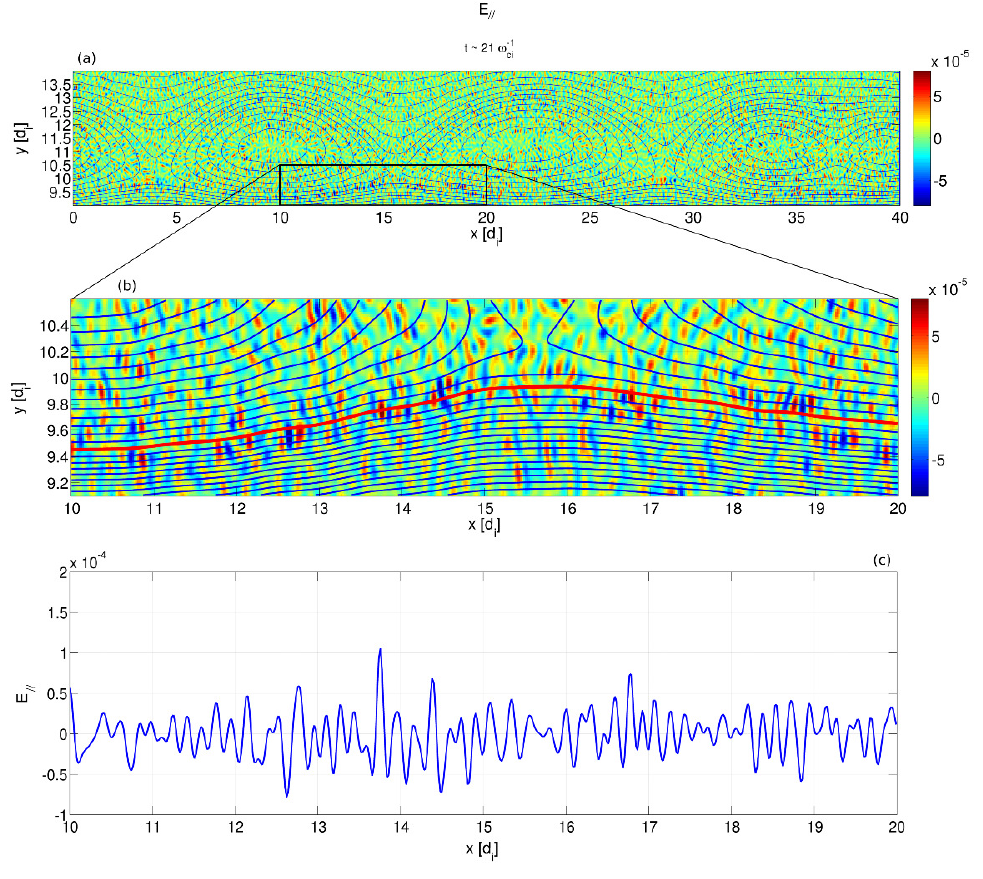} 
    \caption{Parallel component of the electric field (panel a), with zoom of a specific region where a large collection of bipolar fields are observed (panel b). Panel c shows 
    the profile of $E_{\|}$ along the magnetic field line marked in red in panel b. A peak in $E_{\|}$ is seen, e.g., around $x \sim 13.8 \text{ } \unit{d_{i}}$. 
    This bipolar structure can be the signature of an electron hole.} \label{fig:Epar_TOT}
 \end{center}
\end{figure}

\begin{figure}
 \begin{center}
  \includegraphics[scale=0.88]{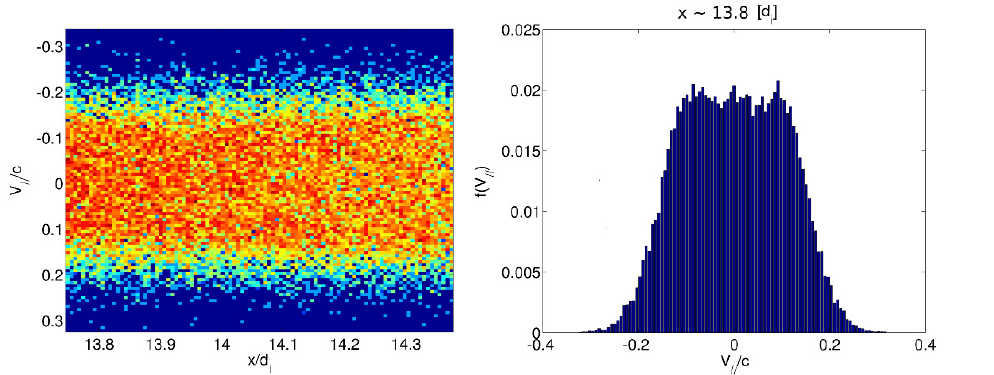} 
    \caption{Electrons phase space (left panel) and velocity distribution of the parallel component (right panel) caught around the peak in $E_{\|}$ seen at $x \sim 13.8 \text{ } \unit{d_i}$. 
    The plots show a flat-top distribution typical signature of a post-instability.} \label{fig:flatopdistr}
 \end{center}
\end{figure}

Figure \ref{fig:recregTOT} shows the electron phase space along two vertical crossing through the lower separatrix in two different regions (i.e . region D and X respectively).
The role of separatrices in  symmetric reconnection has been recently reviewed in \textcite{lapenta2014}, \textcite{Lapentabook}, 
pointing out their importance on particle energization. Two types of energization are indeed experienced in this region: 
 an increase in the bulk velocity and an increase in the thermal energy. Bulk velocity acceleration is due to the presence of a strong 
parallel electric field in the vicinity of the reconnection point, 
which tends to accelerate the electrons inwards toward the X-line following the separatrices. 
In our case, this is confirmed by the parallel electron velocity plotted in figure \ref{fig:VeparPerp}, which shows the electrons to have a 
relevant parallel velocity component in the proximity of the regions X.
These particles will then follow the separatrices to 
the region D, where these two counter-streams ultimately interact with each other leading to some streaming instability. 
The presence of two counter-streams in the D regions is confirmed by the double parallel stream noticeable in figure \ref{fig:recregTOT} and the 
double beams observed in the velocity distributions shown in figure \ref{fig:VparVperp1recreg}. 
Additionally, in analyzing the parallel component of the electric field shown in figure \ref{fig:Epar_TOT}, we notice the presence of some bipolar structures
predominantly along the lower separatrix (panel a). The zoom given in the panel b of the same figure shows these structure to be particularly close to the D regions. 
Such structures are well-known to appear in the symmetric case close to the reconnection sites \cite{lapenta2010,lapenta2014,divin2012}, but they were not previously reported in asymmetric conditions.
The value of $E_{\|}$ on a significant magnetic field line intercepting some of these bipolar structures  (marked in red in panel b) is  
plotted in  panel c of Fig. \ref{fig:Epar_TOT}.
Despite the large noise, it is possible to clearly identify the parallel electric field to be locally stronger (e.g. at $x \sim 13.8 \text{ } \unit{d_i}$).
On the other hand, an increment of the thermal energy cannot be explained by the sole presence of the parallel electric field, 
which can only affect the drift velocity, not the thermal velocity, of electrons. It has been proven that the main cause of this type of energization resides in the formation of 
diverse instabilities 
driven by the chaotic particles motions in this very narrow and trafficked layer \cite{lapenta2014}.
Hence, the simultaneous presence of two parallel counter-beams and the strong bipolar parallel electric field 
can be a signature for the formation of strong 
localized streaming instabilities, which ultimately lead to a predominantly parallel heating. 
A deeper analysis on the particles motion is given in figure \ref{fig:flatopdistr} by showing the phase-space and the velocity distribution of the electrons 
around the peak observed at $x \sim 13.8 \text{ } \unit{d_i}$ in figure \ref{fig:Epar_TOT}.
Even though the phase-space shown in the left panel does not present any direct signature of the linear phase of growth of instabilities, the corresponding
distribution function plotted in the right panel shows the typical flat-top distribution of the post-saturation of streaming instabilities \cite{dieckmann2004,asano2008}.

In concluding this section, let us give a brief discussion on the non-gyrotropies seen in figure \ref{fig:gyrotropyTOT}. 
As expected, regions X show a marked deviation from the pure gyrotropy due to the ongoing reconnection. 
Likewise, regions M show high non-gyrotropic values exactly where the merging points are located. 
Both these effects are observed to persist in time even after the island merging is concluded. In particular, regions M lead to the formation of the strong non-gyrotropic 
spots observable in the center after the islands merge. 

On the other hand, the most interesting outcome in figure \ref{fig:gyrotropyTOT} is the bipolar double layer underlying the entire domain, which  
is also visible in the upper current sheet shown in the last panel. This double structure corresponds to the conjunction of the hot $T_{\perp_{1}}$ and $T_{\perp_{2}}$ layers 
clearly visible in figures \ref{fig:T0perp1} and \ref{fig:T0perp2}, whose magnitude is given by the stronger effect of, respectively, 
either $T_{\perp_{1}}$ ($T_{\perp_{2}}$) 
in the upper (lower) non-gyrotropic structure. A comparison between panels a and b in figure \ref{fig:gyrotropyTOT} shows that this structure  persists in time.
In the D regions, this non-gyrotropic structure is observed to be quadrupolar. This pattern 
is more evident in $V_{\perp_{1}}$  and 
$V_{\perp_{2}}$ of the phase spaces in figure \ref{fig:recregTOT} (left column). By following the $y$ coordinate, we notice that particles at different $y$ coordinates experience a double series of, 
respectively,  acceleration 
in the out-of-plane perpendicular direction or  heating in the in-plane component.
Conversely, particles in regions X (right panel in figure \ref{fig:recregTOT}) experience only a single series of a strong acceleration in the out-of-plane 
perpendicular direction or a 
thermal heating of the in-plane component.
Finally, the analysis of the sequence of motion of this particular quadrupolar layer shows that it is actually composed of two set of bipolar layers. 
The upper set especially appears to be a spin-off of the underlying double structure, and to be drawn upward
during the islands diverging.
Within this region, the conditions are in fact frozen-in and particles are still completely magnetized. 
They hence tend to follow those magnetic field lines raising from the central region, however maintaining their preferential parallel heating.

\section{\label{sec:islands} Islands}

The instability occurring in the lower current sheet soon leads to the formation of different reconnection points, whose outflows in turn cause the emergence of several magnetic islands rapidly 
growing and coalescing with each other. The resulting impact on particle energization is strongly related to this irregular evolution, as seen in 
figures \ref{fig:T0par}, \ref{fig:T0perp1} and \ref{fig:T0perp2}.
These figures show that, after an initial rapid merging where the plasmoids present an internal overall uniform heating, at later stages
the internal hot plasma progressively reduce its temperature inward and becomes surrounded by a growing layer of cold plasma. 
This is due to the faster cooling that the outer plasma experiences over the internal one.
This effect can be explained by the continuous expansion of the new larger islands caused by merging process between two smaller ones. The expansion temporarily 
ceases when the islands coalescence 
no longer occurs and a new stable equilibrium between the internal and external plasma is achieved. After the subsequent merging, the islands start to expand 
again.
The importance of this process is that, while individual mechanisms tend to heat in a specific direction, parallel or perpendicular, we find here that an isotropization mechanism is at work. 

In spite of the global cooling, however, a series of hotter rings surrounding a hot center are constantly seen. This effect shows a  different pattern in every direction, 
suggesting the possibility of localized anisotropic heating mechanisms occurring on some preferential closed lines, which prevent the plasma from cooling down as the rest of the inner island. 

A possible explanation is given by considering the combination of two acceleration mechanisms, namely the betatron and Fermi acceleration. 
These two mechanisms are controlled by the constancy of, respectively, the magnetic moment $\mu$ and the longitudinal invariant $J$ \cite{fermi1949,fu2011}.
These invariants can be  assumed recalling that the fields  vary on  much slower scales than the typical particle response time. 
In the specific case of the increase of the parallel temperature, this temporal variation is thought to be caused by the progressive lateral contraction of the islands 
due to the thrust from the nearest reconnection exhausts.
This mechanism is explained by \textcite{drake2006} via the mirror effect the particles experience during this contraction, which lead them to increase their parallel energy.

\begin{figure}
 \begin{center}
  \includegraphics[scale=0.8]{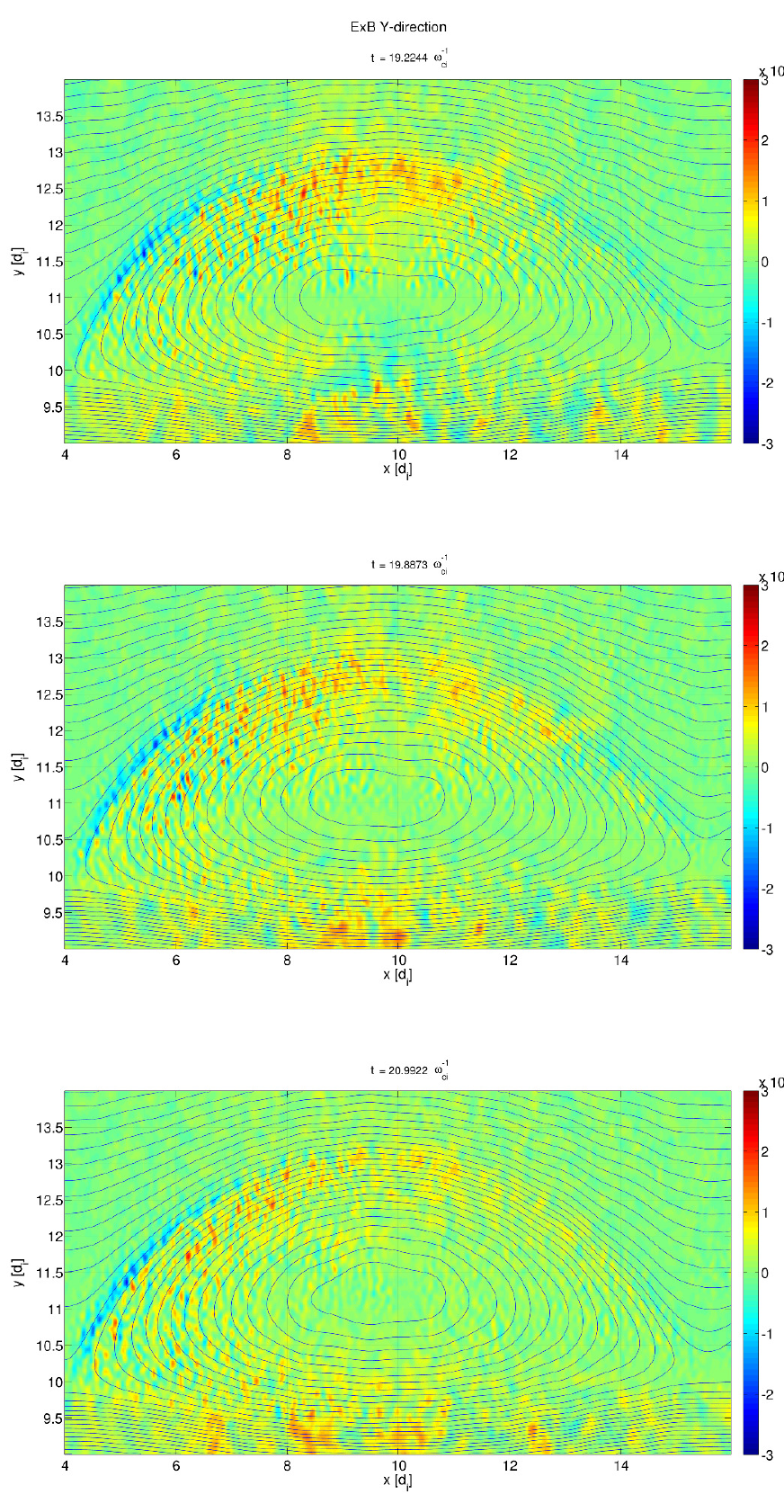} 
    \caption{Magnetic field lines plotted over the Y component of the $\mathbf{E} \times \mathbf{B}$ term at different times. A contraction of the inner islands is not clearly visible.} \label{fig:ExB}
 \end{center}
\end{figure}

However, a clear island contraction is not observed in our case, 
as demonstrated in figure \ref{fig:ExB}, where the $y$ component of the $\mathbf{E} \times \mathbf{B}$ drift is plotted, with superimposed magnetic field lines.

Likewise, the first adiabatic invariant is defined as $\mu \propto \frac{E_{\perp}}{B}$, where $E_{\perp}$ is the perpendicular energy.
In order for $\mu$ to be constant, $E_{\perp}$ must increase together with 
the magnitude of the magnetic field. This mechanism is known as betatron or adiabatic acceleration. 
This effect can explain that inner region of reduced perpendicular temperature observed around the centre of the islands both before and after the merging. 
These regions in fact represent the inflow regions of the M-type points explained earlier. 
Specifically, we know from the symmetric case that the inflowing particles tend to decrease their perpendicular energy 
to compensate the simultaneous decrement of the magnetic field in the same region \cite{egedal2009}, a mechanism that leads to anisotropy in the inflowing region of   reconnection.

Finally, we analyze the localized heated patches visible in the perpendicular components and marked with $1$ and $2$ in figure \ref{fig:T0perp1} (panel d).

\begin{figure}
 \begin{center}
  \includegraphics[scale=0.8]{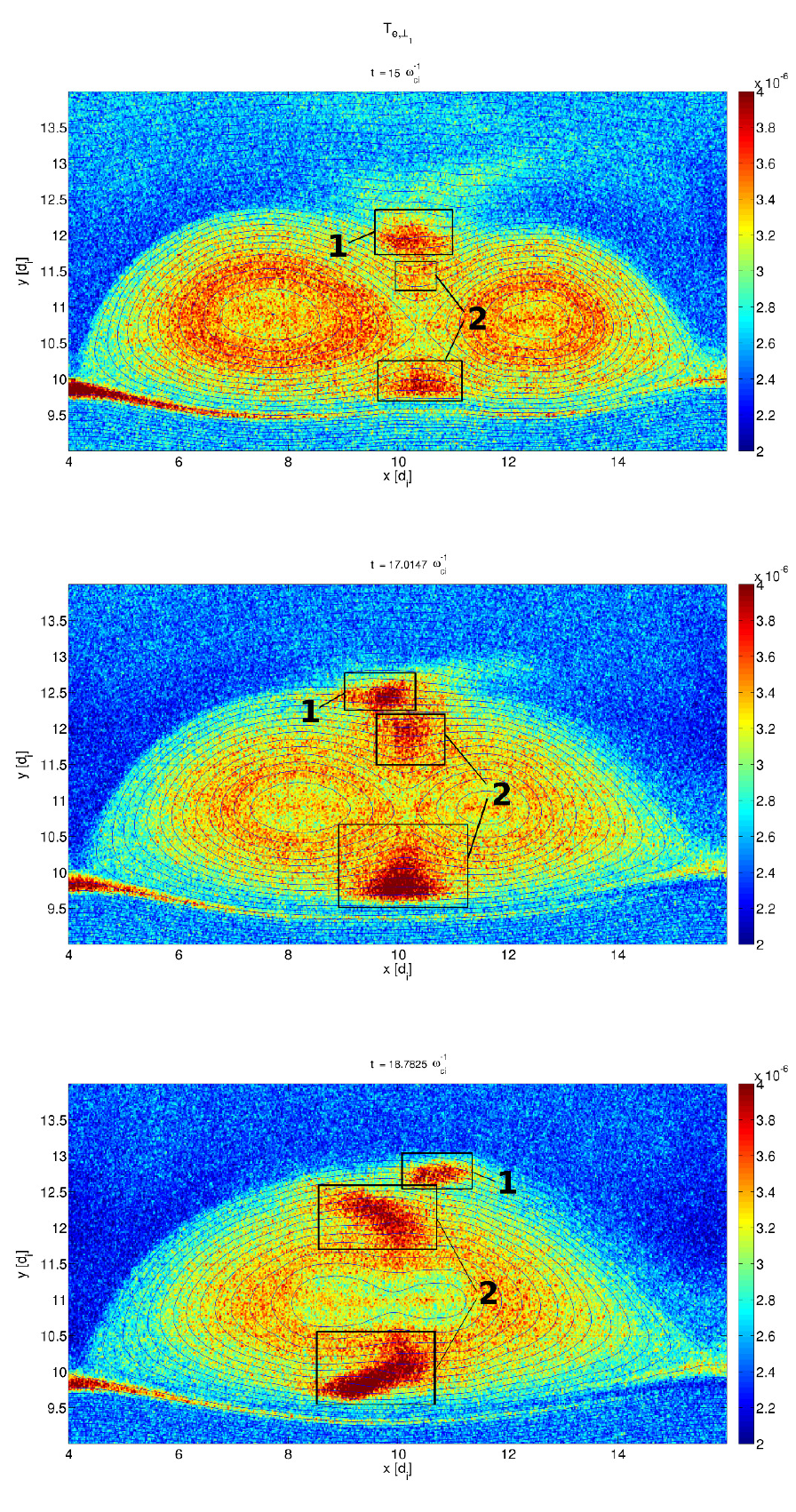} 
    \caption{Evolution of the localized heated patch seen in $T_{\perp_{1}}$ at different times (i.e. $ t \sim 19.2 \text{, } \sim 19.9 \text{ and } \sim 21 \text{ }
    \unit{\omega_{ci}^{-1}}$. This plot shows that the internal patch is indeed the evolution of the reconnection outflows
    occurring between two merging islands. The first outflow is later pushed upward by the second outflows from the further reconnection between two innermost magnetic islands. However, these two 
    flows appear to finally behave differently according to the field line they are shifted on.} \label{fig:Tperp1locheat}
 \end{center}
\end{figure}

$T_{\perp_{1}}$ is presented at different times for the same island
in figure \ref{fig:Tperp1locheat} to show the overall evolution of this strong perpendicular localized 
heating. They are clearly originated by the merging of two islands.

In particular, we can distinguish two patches following a different behavior.
The \emph{patch} marked with $2$ is observed to stream leftwards within the island, and can in turn be split in the twin
streams at the top and the bottom. 
The second \emph{patch}, marked with $1$, instead lies outside the island and is observed to move rightwards toward the X-line 
 following the external field lines bordering the separatrix.  
In figure \ref{fig:Tperp1locheat} we can clearly notice the presence of two subsequent jets flowing out from two consecutive merging. 
The first jet is initially generated by the 
first merging and later pushed upwards by the upcoming second jet (both jets are originated and visible at $x \sim 10.2$). This event finally generates the \emph{patch} marked with $1$. 
The \emph{patch} marked with $2$ is instead originated by the second merging nearly in the same position and is observed to act at replacing the first jet. 
This behavior is further confirmed by observing the twin stream at the bottom.
Both these patches initially tend to flow leftwards. However,
while they are being pushed upwards by following the island expansion, the patch $1$ inverts its motion from left to right 
when it reaches the outer field line, by flowing back toward the X-line along the upper separatrix.
At the end, both of them are soon cooled and absorbed in the inner background plasma. 
These peculiar dynamics can be explained by the difference in velocity of the merging islands. 
As evident from panels b and c in \ref{fig:Tperp1locheat}, the island approaching from right to left moves faster than that approaching from the left.
This can be explained by the different internal momentum of the frozen-in plasma due to the different 
dimension of the two islands.

\section{\label{sec:conclusions} Summary and Conclusions}

We present the results from a fully kinetic simulation of a double periodic box with two differently configured asymmetric layers. The upper layer is initialized with continuous profiles of 
density and magnetic field
as  suggested in \textcite{pritchett2008}. In the lower layer, a strong discontinuity is set up at initialization, resembling the conditions of a
typical Riemann's problem (e.g. \textcite{ferraro1952}). 

The lower layer soon becomes unstable, leading
to the formation of multiple reconnection points with the  formation and coalescence of several magnetic islands.
The centers of the growing islands move upward and all the process
evolves far from the original layer. 
This effect is explained in \textcite{oka2010} as a consequence of the fact that some of the initial reconnection
points are weak, rapidly vanishing and causing a mutual attraction between the two neighboring islands. Once these islands encounter, a reversed 
reconnection event can take place, which has sometime been defined as anti-reconnection \cite{pritchett2008b,oka2010}. 

The main aim of this work is to analyze the electron dynamics resulting from the violent multiple reconnections and subsequent island merging at the lower layer, with particular focus
on the inner islands, the reconnection regions and the separatrices. The electrons heating along the directions parallel and perpendicular to the local magnetic field lines have been taken 
into account more in details (figures \ref{fig:T0par}, \ref{fig:T0perp1} and \ref{fig:T0perp2}).
The process of the island merging is more evident at later times, when the number of islands is reduced and their size larger. 

Islands merge as follow.
After the collapse of an intermediate X-line, they are attracted to each other because of the localized current and finally merge together to
form a larger island. This resulting island in turn expands by progressively cooling down from the outer layer inwards, until the expansion ceases due to the achievement of a temporarily stable condition
between the internal 
and external plasmas. Finally, two neighboring islands start again to feel each other's attractions and eventually coalesce together by repeating the whole cycle on progressively larger scales.

More interesting is however the result obtained by analyzing the presence of anisotropies and agyrotropies. Three main regions have been identified according to their different properties.
Regions marked with X represent the traditional reconnection events occurring horizontally with a vertical particle inflows from the top and the bottom. On the contrary, regions marked with M identify those
reconnection events occurring during the island merging, which are in particular observed to happen between the inner closed magnetic field lines. 

Thanks to this analysis, we were able to understand the origins of the hotter patches visible in the perpendicular components and surrounding the centers of the expanding islands, 
in spite of their continuous cooling.
These patches are generated by the outflows of the reversed reconnection events 
occurring between the merging of the internal closed magnetic fields of two merging islands. Further investigations show them to be indeed two jets from two consecutive reconnection events. The first jet is
soon pushed upwards by the upcoming second jet. However, even though their initial motion appears to be in the same direction, the first jet reverses its motion when the outer magnetic lines
are reached, by now flowing along the separatrices towards the X-line of the nearest region X.

Finally, the D regions identify a third type of region which shows an inverse heating with respect to the regions X. In fact, the presence of a narrow anisotropic layer 
bordering the side with stronger magnetic field is seen in both the X and D regions. Those marked with X show a specific increment of the perpendicular components whereas the D regions 
show an increment of the parallel one. A further analysis on the parallel electron velocity (figure \ref{fig:VeparPerp}) revealed that the islands forming the D regions
diverge. Figure \ref{fig:gyrotropyTOT} shows that the centers of these regions are nearly gyrotropic, indicating the absence of any reconnection events. 
Figure \ref{fig:gyrotropyTOT} shows the presence of a considerable non-gyrotropic bipolar structure surrounding the whole domain along the side with the stronger 
magnetic field and lower density. The same pattern, but reversed, is also observed in the upper current sheet (same figure, panel c).
This bipolar structure forms since the earliest stages and persist along the whole simulation.
From figures \ref{fig:T0par}, \ref{fig:T0perp1} and \ref{fig:T0perp2}, we notice that this structure is formed by the proximity of two narrow layers dominated by, 
respectively, $T_{e,\perp_{2}}$ ($T_{e,\perp_{1}}$) in the outer (inner) layer.
The parallel component is globally absent, becoming relevant only near the D regions.

Interestingly, the bipolar structure turns into a quadrupolar structure in the vicinity of the D regions, where a double alternation of the bipolar layer mentioned above is observed. This quadrupolar structure is not 
seen in the upper current sheet.
By analyzing the sequence of motions,
this secondary double layer is seen to detach from the underlying structure, as the particles were drawn upwards during the islands diverging. 
Within this region conditions are in fact still frozen-in and particles completely magnetized. 
They therefore tend to follow those magnetic field lines raising from the central region by maintaining their preferential parallel heating.

The results presented provide important observational hints into what the  Magnetospheric Multiscale  mission (MMS)  by NASA is likely to observe when passing regions of multiple reconnection site in the Earth magnetopause.

In particular, this work can contribute to the understanding of the formation and evolution of magnetic flux ropes, the three dimensional extension of the 2.5D magnetic islands  analyzed here. Remarkable signatures have been evidenced
by these simulations, including the strong anisotropy and agyrotropy grown on specific sides of the system, as well as the interesting vertical reconnection events between merging islands, 
whose direct observation and validation will now be possible thanks to the unprecedented time and space resolution designed for the MMS mission.

\begin{acknowledgments}
 
The present work is supported by the NASA MMS Grant
NNX08AO84G. Additional support for the KULeuven
team is provided by the European Commission DEEP-ER project, by the Onderzoekfonds KU Leuven (Research Fund KU Leuven) and 
by the Interuniversity Attraction Poles Programme of the Belgian Science Policy Office (IAP P7/08 CHARM). 

The simulations were conducted on the Pleiades supercomputer of the NASA Advanced Supercomputing Division (NAS), on the Discover supercomputer of the NASA Center for Climate Simulation (NCCS) 
on the computational resources provided by the PRACE Tier-0 2013091928  (SuperMUC supercomputer) and 
on the Flemish Supercomputing Center (VSC-VIC3) and on the SuperMUC supercomputer thanks to a PRACE Tier-0 research infrastructure grant.
 
\end{acknowledgments}

%
%

%

\end{document}